\documentclass[%
 aip,
 jmp,%
 amsmath,amssymb,
 reprint,%
]{revtex4-1}

\usepackage{graphicx}
\usepackage{dcolumn}
\usepackage{bm}
\usepackage{mathrsfs}
\usepackage{amsthm}
\usepackage{easybmat}

 \newcommand{\U}{\mathcal U}
 \newcommand{\ob}[1]{\mathscr{#1}}
\newcommand{\schrodinger}{Schr\"odinger}
\newcommand{\exps}[1]{\operatorname e^{#1}}
\newcommand{\braket}[2]{\left < {#1 \,| \, #2} \right >}
  \newcommand{\braketl}[2]{\left < \right. {#1 \,| \, #2} \left > \right.}

\begin{document}
\title[Time dependent quantum generators for the Galilei group]{Time dependent quantum generators for the Galilei group}

\author{G. Filippelli}
 \altaffiliation[Also at ]{Istituto Nazionale di Fisica Nucleare, Gruppo collegato di Cosenza\\Electronic mail: gianluigi.filippelli@gmail.com}
\affiliation{ 
Dipartimento di Matematica, Universit\`a della Calabria
}%
\begin{abstract}
In 1995 Doebner and Mann introduced an approach to the ray representations of the Galilei group in ($1+1$)-dimensions, giving rise to quantum generators with an explicit dependence on time. Recently (2004) Wawrzycki proposed a generalization of Bargmann's theory: in his paper he introduce phase exponents that are explicitely dependent by 4-space point. In order to find applications of such generalization, we extend the approach of Doebner and Mann to higher dimensions: as a result, we determine the generators of the ray representation in ($2+1$) and ($3+1$) dimensions. The differences of the outcoming formal apparatus with respect to the smaller dimension case are established.
\end{abstract}

\maketitle

\section{\label{sec:intro}Introduction}
In 2004 Wawrzycki proposes a generalization of Bargmann's theory \cite{Waw04}, introducing the phase exponents that explicitelly depends on time, for non-relativistic groups, and phase exponents that explicitelly depends on 4-space point, for relativistic groups.

On the other hand, in 1995, Doebner and Mann \cite{DoebnerMann95}, studying Gelilei group in $(1+1)$ dimensions, calculated the generators group's representations that explicitelly depends on time, while for Galilei group in $(2+1)$ dimensions they founded the generators group's representations just calculated by Bose \cite{Bose95} and Grigore \cite{Grigore96}.

At this point, in order to generalize Doebner and Mann work about higher dimensions, in Section \ref{sec:extension} we calculate the generators of the Galilei group's representations and we show that they explicitely depends on time. These representations are valid for $(2+1)$ and $(3+1)$ dimensions. At the end, from these representations, we calculate the ray representations of Galilei group, and we show that they explicitely depends on time.

First of all, we introduce the ray representations of a given Lie group (Section \ref{sec:rayrepresentations}), and we briefly remind the application of the Bargmann's theory \cite{Bargmann54} to Galilei group (Section \ref{sec:galilei}). In the Section \ref{sec:extension} we find the time dependance of the ray representations of galilean group in $(3+1)$ and $(2+1)$ dimensions, with phase exponents that depend on time, and we propose a physical interpretation either for it, either for Bargmann's phase exponent.

\subsection{\label{sec:rayrepresentations}Ray representations}
The vector ray notion, introduced by Weyl \cite{Weyl31}, can be extended also to the operators, in particular to unitary operators. According to Wigner's theorem \cite{Wigner59, Ulf63, Bargmann64}, every symmetry's transformation $T_r$ can be represented by an unitary (or anti-unitary) operator, which is unique up at a phase factor, id est $T_r$ can be represented by an unitary operator ray defined by:
 \begin{equation}\label{eqr-operatorray}
  \U_r = \{\exps{i \theta} U_r, \theta \in {\mathbb R}\}.
 \end{equation}
So, $\U_e$ is the ray containing the identity operator and $\U^{-1}_r$ is the inverse of $\U_r$, i.e. the ray containing the operators $U^{-1}_r$ inverse of $U_r \in \U_r$,
 \begin{displaymath}
  \U^{-1}_r = \{V_r, \text{ such that } V_r = U_r^{-1}, U_r \in \U_r\}
 \end{displaymath}
then $\U_r \cdot \U^{-1}_r = \U^{-1}_r \cdot \U_r = \U_e$.

Now, let $G$ be a symmetry group with elements $r, s, t, \ldots$; then, for a given choice of the unitary operator $U_r \in \U_r$ representing the elements of $G$, in general the following composition law holds:
 \[U_r U_s = \omega (r,s) U_{rs}\]
that we can rewrite for the operator rays as:
 \[\U_r \cdot \U_s = \U_{rs}\]
where $\omega (r,s)$ is a factor of modulus 1 $(|\omega (r,s)|=1)$, $r,s \in \ob{N}_0 \subset G$, and $\ob{N}_0$ is a neighbourhood of $G$ identity.

The correspondence $r \rightarrow \U_r$ realizes a {\bf ray representation} (or {\em projective representation}) of $G$. It is equivalent to an usual unitary representation of $G$ if a correspondence $r \rightarrow U_r \in \U_r$ exists such that $\omega (r,s) \equiv 1$, $\forall r,s \in G$.

In order to classify the ray representations of a given Lie group $G$, it is sufficient classify the phase factor equivalence classes; indeed, for a different admissible set of representatives\footnote{Let $\U_r$ be a continuous ray representations of a group $G$. For all $r$ in a suitably chosen neighborhood $\ob{N}_0$ of the identity $e$ of $G$, one may select a strongly continuous set of representatives $U_r \in \U_r$.

A set of such representatives operator will be called an {\em admissible} set of representatives. \cite{Bargmann54}} $U'_r = \phi (r) U_r$, we obtain
 \begin{displaymath}
  U'_r \cdot U'_s = \omega' (r,s) U'_{rs}
 \end{displaymath}
where
 \begin{equation}\label{eqr-factorequivalence}
  \omega' (r,s) = \omega (r,s) \frac{\phi(r) \phi(s)}{\phi(rs)}.
 \end{equation}
But it is more advantageous to replace (local) factors with (local) exponents by setting $\omega (r,s) = \exps{i \xi(r,s)}$. So, a phase exponent of a group $G$ is a real valued continuous function $\xi (r,s)$ which is defined for all $r$, $s$ in $G$ and which satisfies the relations \cite{Bargmann54}:
 \begin{subequations}\label{eqr-phaseexponent}
  \begin{align}
   \xi (e,e) = & \, 0\label{eqr-phaseidentity}\\
   \xi (r,s) + \xi (rs, t) = & \, \xi (s,t) + \xi (r, st)\label{eqr-phaseassociativity}
  \end{align}
 \end{subequations}
For every phase exponent defined on the group (or on a neighbourhood), a function called {\em infinitesimal exponent} $\Xi$ defined on the Lie algebra of the group exists, that is in one-o-one linear correspondence with phase exponent:
 \begin{eqnarray}\label{eqr-infinitesimalexponent}
  \Xi (a,b) &=& \lim_{\tau \rightarrow 0} \tau^{-2} \left ( \xi ((\tau a)(\tau b), (\tau a)^{-1} (\tau b)^{-1}) + \right .\nonumber\\
  && + \left . \xi (\tau a, \tau b) + \xi ((\tau a)^{-1}, (\tau b)^{-1}) \right ).
 \end{eqnarray}

\section{\label{sec:galilei}The Galilei's group}
Now, we remind the study about the Galilei's group, its Lie algebra and ray representations in $(3+1)$-dimensions (Section \ref{sec:galileibargmann}), $(2+1)$-dimensions (Section \ref{sec:galileibosegrigore}) and $(1+1)$-dimensions (Section \ref{sec:galileidoebnermann}). In the last case we show how Doebner and Mann \cite{DoebnerMann95} calculate the generators of Galilean group, which turn out to depend on time, and we propose the Galilei's group's ray representations that depend on time.

\subsection{\label{sec:galileibargmann}The Galilei's group in $(3+1)$-dimensions}
The Galilean group is constituted by all space-time transformations from an inertial reference frame to an other one. The most general Galilean transformation of the Galilean group $G$ is:
 \begin{subequations}
  \begin{equation}\label{eqGG-galileitransformation01}
   \left \{
   \begin{aligned}
    x' & = \, Wx + vt + u\\
    t' & = \, t + \eta
   \end{aligned} \right.
  \end{equation}
where $x'$, $x$ are spatial vectors, $v$ is the relative velocity, $u$ is a space translation, $t$ is time and $\eta$ a time translation, with $W$ an orthogonal transformation (e.g. rotation).

In order to classify the ray representations of Galilei's group, we can represent, following Bargmann \cite{Bargmann54}, the generic element $r$ of the Galilean group $G$ as:
  \begin{equation}\label{eqGG-galileitransformation02}
   r = (W_r, \eta_r, v_r, u_r).
  \end{equation}
 \end{subequations}
So, the group multiplication is given by
 \begin{eqnarray}
  rs &=& (W_r, \eta_r, v_r, u_r) \cdot (W_s, \eta_s, v_s, u_s) =\nonumber\\
     &=& (W_r W_s, \eta_r + \eta_s, W_r v_s + v_r, W_r u_s + u_r + \eta_s v_r).\label{eqGG-galileiproduct}
 \end{eqnarray}
Now, to classify the ray representations of $G$, we first must describe the algebra of Galilei's group: algebra standard basis is constituted by $a_{ij}$, anti-symmetric $3 \times 3$ matrix where only elements $ij$ are no-null; $b_i$, the translations generator; $d_i$, the pure galilean transformations generator ($1 \leqslant i \leqslant n$, $1 \leqslant j \leqslant n$, where $n$ is the space dimension); $f$, the time translations generator. The generators algebra is given by:
 \begin{subequations}\label{eqGG-algebracommutators}
  \begin{align}
   [a_{ij}, a_{kl}] = & \, \delta_{jk}a_{il} - \delta_{ik} a_{jl} + \delta_{il} a_{jk} - \delta_{jl} a_{ik} \label{eqGG-rotationcommutaor}\\
   [a_{ij}, b_k] = & \, \delta_{jk} b_i - \delta_{ik} b_j; \qquad [b_i, b_j] = 0 \label{eqGG-translationcommutator}\\
   [a_{ij}, d_k] = & \, \delta_{jk} d_i - \delta_{ik} d_j; \qquad [d_i, d_j] = 0; \qquad [d_i, b_j] = 0 \label{eqGG-galileicommutator} \\
   [a_{ij}, f] = & \, 0; \qquad [b_k, f] = 0; \qquad [d_k, f] = b_k \label{eqGG-timecommutator}
  \end{align}
 \end{subequations}
At this point we can calculate all infinitesimal exponent of the Galilei's group. The only no-null exponent is:
 \begin{equation}\label{eqGG-galileiinfinitesimalexponent}
  \Xi (b_i, d_k) = - \Xi (d_k, b_i) = \gamma \delta_{ik}.
 \end{equation}
The corresponding phase exponent $\xi$ is a multiple of the function $\xi_0$:
 \begin{eqnarray}
  \xi (r,s) &=& \gamma \xi_0 (r,s) = \frac{1}{2} \gamma \left ( \braket{u_r}{W_r v_s} + \right.\nonumber\\
   && - \left. \braket{v_r}{W_r u_s} + \eta_s \braket{v_r}{W_r v_s} \right ),\label{eqGG-galileiexponent3D}
 \end{eqnarray}
by a multiplicative factor $\gamma$, which is interpreted as the mass of a free particle\footnote{From the study of the full Galilean group in $(3+1)$ dimensions, it is possible to obtain the Bargmann's super-selection rules \cite{Brennich70}:

1) To different masses there correspond inequivalent multipliers and hence inequivalent ray representations;

2) ${\rm SO} (3)$ has two kind of inequivalent multipliers; one for integer spin (unitary representations), a second kind for semi-integer spin (ray representations) \cite{Wigner39, GKWigner68}}.

At this point Bargmann determined the Galilean ray representations \cite{Bargmann54}:
 \begin{eqnarray}
  \phi' (p) &=& U_r \phi (r^{-1}p) =\nonumber\\
            &=& \exps{-i (\braket{p}{u} - \frac{\eta}{2 \gamma} \braket{p}{p} + \frac{\eta}{2}\gamma \braket{v}{v} - \gamma \braket{u}{v})} \phi (r^{-1}p)\label{eqGG-galilei3Drayrep}
 \end{eqnarray}
where the functions $\phi (p)$ are the wave functions in the {\em Heisenberg representation}
 \footnote{Let be $\psi(x,t)$ are the solutions of {\schrodinger} equation, then:
  \begin{displaymath}
   \psi (x,t) \sim (2\pi)^{-\frac{3}{3}} \int \phi (p) \exps{-i \frac{1}{2 \gamma} \braket{p}{p}t - \braket{p}{x} \operatorname d^3 p}
  \end{displaymath}} ($\phi (p) \in \mathcal L_2 ( \mathbb{R}^2 , p)$), and $p$ being the linear momentum.

\subsection{\label{sec:galileibosegrigore}The Galilei's group in $(2+1)$-dimensions}
Now we can study the Galilean group in lower dimensions. By the results obtained by Bargmann \cite{Bargmann54} about the pseudo-orthogonal groups, we must expect the emergence of new phase exponents in $(2+1)$ dimensions. Indeed, in this case, there are two non-equivalent non-trivial phase exponents \cite{Bose95, Grigore96} besides $\xi_0$ in (\ref{eqGG-galileiexponent3D}):
 \begin{subequations}\label{eqGG-galilei2Dexponents}
  \begin{align}
   \xi_1 (r,s) = & \, \frac{1}{2} ( v_r \wedge W_r v_s ) \label{eqGG-galileiexponent2D01}\\
   \xi_2 (r,s) = & \, \theta_r \eta_s - \theta_s \eta_r \label{eqGG-galileiexponent2D02}
  \end{align}
 \end{subequations}
where $(u \wedge v) = u_1 v_2 - v_1 u_2$, with $u$, $v$ two-dimensional vectors.

All inequivalence classes are multiple of (\ref{eqGG-galilei2Dexponents}) with multiplicative factors $\lambda$ and $S$ respectively. In correspondence with different values of the three multiplicative factors $\gamma$, $\lambda$, $S$, we can have non-equivalent ray representations of the Galilean group. In particular, we are interested in the ray representations with $S=0$; so we have the two following cases:
 \begin{align}
  (U(W, \eta, v, u) f)(p) = & \, \exps{i \left ( \braket{u}{p} + \frac{\gamma}{2} \braket{u}{v} + \frac{\eta}{2 \gamma} \braket{p}{p} - \frac{\lambda}{2 \gamma} ( v \wedge p) + s \theta \right )}\nonumber\\& f(W^{-1} (p + \gamma v)) \label{eqGG-galilei2Dschrodingernonabelianrep}\\
  (U(W, \eta, v, u) f)(p) = & \, \exps{i \left ( \braket{u}{p} + \frac{\gamma}{2} \braket{u}{v} + \frac{\eta}{2 \gamma} \braket{p}{p} \right )}\nonumber\\& s(h) f \left ( W^{-1} (p + \gamma v) \right ) \label{eqGG-galilei2Dschrodingerrep}
 \end{align}
where (\ref{eqGG-galilei2Dschrodingernonabelianrep}) corresponds to a localizable {\schrodinger} system for which the {\em boost} representation is not abelian, and (\ref{eqGG-galilei2Dschrodingerrep}) describes a {\schrodinger} non-relativistic particle, where $s(h)$ denotes an irreducible representation of the sub-group $\{(W, 0, v, 0)\}$, and the functions $f(p) \in \mathcal L_2 (X_r, p)$, where $X_r = \{(p_0,p)\}$, with $\braket{p}{p} = r^2$, is an orbit.

\subsection{\label{sec:galileidoebnermann}The Galilei's group in $(1+1)$-dimensions}
According to Doebner and Mann \cite{DoebnerMann95}, we write the generic element of the Galilean group in $(1+1)$ dimensions as
 \begin{equation}\nonumber
  r = (u_r, v_r, \eta_r)
 \end{equation}
where $u_r$ is a space translation, $v_r$ a velocity translation, $\eta_r$ a time translation. The corresponding phase exponent is:
 \begin{eqnarray}
  \xi_\eta (r,s) = & \frac{a_1}{2} (a_r v_s - a_s v_r + \eta_r v_r v_s) +\nonumber\\
  & à+ \frac{a_2}{2} (u_r \eta_s - u_s \eta_r - \eta_r \eta_s v_r)\label{eqGG-galileiexponent1D}
 \end{eqnarray}
with $a_{1,2}$ are real numbers.

To introduce a time dependance in Hilbert space we use a kind of Hisenberg picture. For any self-adjoint operators $R (X)$ we set \cite{Messiah75}:
 \begin{equation}\label{eqGG-heisenbergpicture}
  \frac{\operatorname d}{\operatorname d t} R_t(X) = K R_t ([H,X])
 \end{equation}
with initial condition
 \[R_{t=0} (X) = R(X)\]
(where $K$ is a complex constant, $H$ time translations generator, $R_t (X)$ is the time depending representation of the generator $X$ and $\frac{\partial X}{\partial t}=0$) we can calculate generators of Galilean group:
 \begin{eqnarray}
  R_t (H) &=& - \frac{\hbar^2}{2 m} \partial^2_x + fx + V_0, \qquad R_t (P) = i \hbar \partial_x - ft,\nonumber\\
  R_t (N) &=& mx - i \hbar t \partial_x - \frac{1}{2} f t^2\label{eqGG-galileiinfinitesimaltime}
 \end{eqnarray}
where $H$ is the time translations generator, $P$ the space translations generator, $N$ the boost translation generator, $\hbar$ the Planck constant, $m = a_1$ the mass particle, $f = a_2$ an external force, $V_0 = \frac{a_3}{2 a_1}$\footnote{where $a_3$ is a real number connected to the Casimir element
 \begin{displaymath}
  C_3 = 2 H Z_1 - 2 K Z_2 - P^2
 \end{displaymath}
where $H$, $K$, $P$ are time translations, nonrelativistic boosts and space translations generators, and $Z_1$ and $Z_2$ are the two central elements}.

\section{\label{sec:extension}Extension of Doebner and Mann calculation}
Now we extend the Doebner and Mann approach to determine the time depending Galilean generators in $(2+1)$ and $(3+1)$ dimensions; by using these generators, we derive the unitary time depending representations for the Galilean group. As a result, besides (\ref{eqGG-galileiexponent3D}), we find ray representations with a phase exponent which explicitly depends on time.

Now, applying (\ref{eqGG-heisenbergpicture}) to generators of Galilei's group in $(2+1)$- and $(3+1)$-dimensions\footnote{The infinitesimal generators of (\ref{eqGG-galilei2Dschrodingernonabelianrep}) are:
 \begin{align*}
  H = & \, \frac{1}{2 \gamma} \left ( p_1^2+p_2^2 \right ) & N_1 = & \, \gamma \frac{\partial}{\partial p_1} + i \frac{\lambda}{2 \gamma} p_2\\
  M = & \, is + p_2 \frac{\partial}{\partial p_1} - p_1 \frac{\partial}{\partial p_2} & N_2 = \, & \gamma \frac{\partial}{\partial p_2} - i \frac{\lambda}{2 \gamma} p_1\\
  P_i = \, & p_i \, (i=1,2) &\,&\,
 \end{align*}
The infinitesimal generators of (\ref{eqGG-galilei2Dschrodingerrep}) are:
 \begin{eqnarray*}
  H &=& \frac{1}{2 \gamma} \left ( p_1^2+p_2^2 \right ), \quad P_i = p_i, \quad M = is + p_2 \frac{\partial}{\partial p_1} - p_1 \frac{\partial}{\partial p_2},\\N_i &=& \gamma \frac{\partial}{\partial p_i} (i=1,2)
 \end{eqnarray*}
These ones are equivalent also for (3+1)-dimensions case.}, we find that $R_t (H) = R(H)$, $R_t (P_i) = R (P_i)$, $R_t (M) = R(M)$, while for $N_i$ generators we find:
 \begin{align*}
  R_t (N_1) = & -i p_1 t + \gamma \frac{\partial}{\partial p_1} - i \frac{\lambda}{2 \gamma} p_2,\\
  R_t (N_2) = & -i p_2 t + \gamma \frac{\partial}{\partial p_2} - i \frac{\lambda}{2 \gamma} p_1
 \end{align*}
for representation (\ref{eqGG-galilei2Dschrodingernonabelianrep}), and we find:
 \begin{displaymath}
  R_t (N_i) = -i p_i t + \gamma \frac{\partial}{\partial p_i}
 \end{displaymath}
for representation (\ref{eqGG-galilei2Dschrodingerrep}).

Now, for $v = (1,0,v,0) \in G$, the corresponding unitary operator time depending will be
 \begin{eqnarray*}
  \left ( 1 - i \braket{p}{v}t - i \frac{\lambda}{2 \gamma} (v \wedge p) + \gamma v_1 \frac{\partial}{\partial p_1} + \gamma v_2 \frac{\partial}{\partial p_2} \right ) f(p) \simeq\\
  \left ( 1 - i \braket{p}{v} t - i \frac{\lambda}{2 \gamma} (v \wedge p) \right ) \cdot\\
  \cdot \left ( 1 + \gamma v_1 \frac{\partial}{\partial p_1} + \gamma \frac{\partial}{\partial p_2} \right ) f (p) \simeq\\
  \exps{-i \braket{p}{v}t - i \frac{\lambda}{2 \gamma} (v \wedge p)} f (p+\gamma v) = (U_t (v) f) (p)
 \end{eqnarray*}
for representation (\ref{eqGG-galilei2Dschrodingernonabelianrep}), and
 \begin{displaymath}
  (U_t (v) f) (p) = \exps{-i \braket{p}{v}t} f (p+\gamma v)
 \end{displaymath}
for representation (\ref{eqGG-galilei2Dschrodingerrep}). So the representation of Galilean group, time depending, in $(2+1)$ (for $\gamma\not=0$, $\lambda\not=0$, $S=0$, and for $\gamma\not=0$, $\lambda=S=0$) and $(3+1)$ dimensions is
 \begin{equation}\label{eqGG-timerepresentation}
  (U_t (r) f)(p) = \exps{-i \braket{p}{v_r}t} (U(r) f)(p).
 \end{equation}
So, $\forall r \in G$, the corresponding $U_t (r)$ time depending unitary operator is given by the unitary operator $U (r)$ of the ray representation up to a time depending phase factor, $\exps{-i\braket{p}{v_r}t}$.

Now we can see whether the introduction of the phase $\exps{i\braket{p}{v}t}$ in the ray representation of Galilean group generates a further projective phase factor. To this purpose we calculate:
 \begin{align*}
  \left ( (U_t (r) U_t (s)) f \right ) (p) = & \, \exps{-i \braket{p}{v_r}t} U_t (s) ((U(r) f) (p))\\
  = & \, \exps{-i \braket{p}{v_r}t} \exps{-i \braketl{W_r^{-1}(p + \gamma v_r)}{v_s}t}\\
  & \omega(r,s) (U(rs) f) (p).
 \end{align*}
On the other hand
 \begin{align*}
  (U_t (rs) f) (p) = & \exps{-i \braket{p}{v_{rs}}t} (U (rs) f) (p) =\\
  = & \exps{-i \braket{p}{v_r}t} \exps{-i \braket{p}{W_r v_s}t} (U (rs) f) (p).
 \end{align*}
Comparing these two equations we can find that
 \begin{subequations}
  \begin{equation}\label{eqGG-galileitimeunitaryrep}
   \left ( (U_t (r) U_t (s)) f \right ) (p) = \phi (r,s,t) \, \omega (r,s) \, (U_t (rs) f) (p)
  \end{equation}
where $\omega (r,s)$ is the usual projective phase factor, while $\phi (r,s,t)$ is defined by:
  \begin{equation}\label{eqGG-galileitimemultiplier}
   \phi (r,s,t) = \exps{-i \gamma \braket{v_r}{W_r v_s} t} = \exps{i \xi_t (r,s)}
  \end{equation}
with
  \begin{equation}\label{eqGG-galileitimeexponent}
   \xi_t (r,s) = - \gamma \braket{v_r}{W_r v_s} t = -\gamma \xi_{0, t} (r,s)
  \end{equation}
 \end{subequations}
(\ref{eqGG-galileitimeexponent}) is a bilinear continuous function in $r$, $s$ coordinates which satisfies equations (\ref{eqr-phaseexponent}), like every projective phase exponent.

\subsection{\label{sec:interpretation}Physical interpretation of the phase exponents}
The phase exponent (\ref{eqGG-galileiexponent3D}) of galilean group
 \begin{align*}
  \xi (r,s) = & \gamma \xi_0 (r,s) =\\
  = & \frac{1}{2} \gamma \left ( \braket{u_r}{W_r v_s} - \braket{v_r}{W_r u_s} + \eta_s \braket{v_r}{W_r v_s} \right )
 \end{align*}
has the physical dimensions of a action, and so it can be interpreted like the action of the particle in the frame $rs$. Indeed, set $r = (1,0,0,u)$, $s = (0,0,v,0)$, then $\xi (r,s) = \frac{\gamma}{2} \braket{u}{v}$, that it has the dimension of an action, and $rs = (1,0,v,u)$ is the Galilean transformation that relates an inertial frame system to another with relative velocity $v$ and origin of axes translated by $u$.\\
About the phase exponent (\ref{eqGG-galileitimeexponent}), we propose the following interpretation.\\
Let be $W_r v_s$ velocity of $\Sigma_s$ in $\Sigma_r$; so $\braket{v_r}{W_r v_s}$ is the velocity of $\Sigma_s$ along the motion direction of $\Sigma_r$. So $\xi_t (r,s)$ is the contribution of the two coordinate systems $\Sigma_r$, $\Sigma_s$ to the total action of the particle.\\
For example: let be $r = (1,0,v_r,0)$, $s = (1,0,v_s,0)$ two elements of galilean group. Then
 \begin{displaymath}
  (U_t (r) U_t (s) f)(p) = \exps{-i \gamma \braket{v_r}{v_s}t} (U_t (rs)f) (p).
 \end{displaymath}
And the contribution of the two coordinate systems to the total action of the particle is:
 \begin{displaymath}
   - \gamma \braket{v_r}{v_s}t.
 \end{displaymath}
{\bf In conclusion:} In this work, we brief remind the theory of phase exponents of ray representation of Lie groups and the application on galilean group in $(3+1)$, $(2+1)$ and $(1+1)$ dimensions. Finally, we find the time dependance of the ray representations of galilean group (\ref{eqGG-timerepresentation}): the action of this representation on physical state $f(p)$ is given by:
 \begin{equation}
  f' (p,t) = \exps{-i \braket{p}{v_r}t} \left ( U(r) f \right ) (p)
 \end{equation}
For example, if we use (\ref{eqGG-timerepresentation}) on (\ref{eqGG-galilei3Drayrep}) we obtain:
 \begin{align}
  \phi' (p,t) =\nonumber\\
  \exps{-i \braket{p}{v_r} t} \exps{-i (\braket{p}{u_r} - \frac{\eta_r}{2 \gamma} \braket{p}{p} + \frac{\eta_r}{2}\gamma \braket{v_r}{v_r} - \gamma \braket{u_r}{v_r})} \phi (r^{-1}p)\label{eqGG-lastequation}
 \end{align}
The phase exponents in (\ref{eqGG-lastequation}) represents the total action of the particle in $\Sigma_{rs}$ frame.

\begin{acknowledgments}
I wish to acknowledge Giuseppe Nisitc\'o, at Mathematics Department at Universit\'a della Calabria, for support and advice during reserach work, and Riccardo Adami, at Mathematics Department at Universit\'a Bicocca in Milano, and Stefano Sandrelli, at Brera's Astronomical Observatory in Milano, for the credit in my abilities.
\end{acknowledgments}

\nocite{*}

\begin{thebibliography}{10}%
	\makeatletter
	\providecommand \@ifxundefined [1]{%
		\ifx #1\undefined \expandafter \@firstoftwo
		\else \expandafter \@secondoftwo
		\fi
	}%
	\providecommand \@ifnum [1]{%
		\ifnum #1\expandafter \@firstoftwo
		\else \expandafter \@secondoftwo
		\fi
	}%
	\providecommand \enquote [1]{``#1''}%
	\providecommand \bibnamefont  [1]{#1}%
	\providecommand \bibfnamefont [1]{#1}%
	\providecommand \citenamefont [1]{#1}%
	\providecommand\href[0]{\@sanitize\@href}%
	\providecommand\@href[1]{\endgroup\@@startlink{#1}\endgroup\@@href}%
	\providecommand\@@href[1]{#1\@@endlink}%
	\providecommand \@sanitize [0]{\begingroup\catcode`\&12\catcode`\#12\relax}%
	\@ifxundefined \pdfoutput {\@firstoftwo}{%
		\@ifnum{\z@=\pdfoutput}{\@firstoftwo}{\@secondoftwo}%
	}{%
	\providecommand\@@startlink[1]{\leavevmode}%
	\providecommand\@@endlink[0]{}%
}{%
\providecommand\@@startlink[1]{%
	\leavevmode
	\pdfstartlink
	attr{/Border[0 0 1 ]/H/I/C[0 1 1]}%
	user{/Subtype/Link/A<</Type/Action/S/URI/URI(#1)>>}%
	\relax
}%
\providecommand\@@endlink[0]{\pdfendlink}%
}%
\providecommand \url  [0]{\begingroup\@sanitize \@url }%
\providecommand \@url [1]{\endgroup\@href {#1}{\urlprefix}}%
\providecommand \urlprefix [0]{URL }%
\providecommand \Eprint[0]{\href }%
\@ifxundefined \urlstyle {%
	\providecommand \doi [1]{doi:\discretionary{}{}{}#1}%
}{%
\providecommand \doi [0]{doi:\discretionary{}{}{}\begingroup
	\urlstyle{rm}\Url }%
}%
\providecommand \doibase [0]{http://dx.doi.org/}%
\providecommand \Doi[1]{\href{\doibase#1}}%
\providecommand \selectlanguage [0]{\@gobble}%
\providecommand \bibinfo [0]{\@secondoftwo}%
\providecommand \bibfield [0]{\@secondoftwo}%
\providecommand \translation [1]{[#1]}%
\providecommand \BibitemOpen[0]{}%
\providecommand \bibitemStop [0]{}%
\providecommand \bibitemNoStop [0]{.\EOS\space}%
\providecommand \EOS [0]{\spacefactor3000\relax}%
\providecommand \BibitemShut [1]{\csname bibitem#1\endcsname}%
\bibitem{Waw04}%
\BibitemOpen
\bibfield{author}{%
	\bibinfo {author} {\bibfnamefont{Jaroslaw}\ \bibnamefont{Wawrzycki}},\ }%
\bibfield{journal}{%
	\Doi{10.1007/s00220-004-1141-4}{\bibinfo {journal} {Communications in
			Mathematical Physics}}\ }%
\textbf{\bibinfo {volume} {250}},\ \bibinfo {pages} {215} (\bibinfo {year}
{2004})\BibitemShut{NoStop}%
\bibitem{DoebnerMann95}%
\BibitemOpen
\bibfield{author}{%
	\bibinfo {author} {\bibfnamefont{H.-J.~Mann}\ \bibnamefont{H.-D.~Doebner}},\
}%
\bibfield{journal}{%
	\Doi{10.1063/1.531026}{\bibinfo {journal} {Journal of Mathematical Physics}}\
}%
\textbf{\bibinfo {volume} {36}},\ \bibinfo {pages} {3210} (\bibinfo {year}
{1995})\BibitemShut{NoStop}%
\bibitem{Bose95}%
\BibitemOpen
\bibfield{author}{%
	\bibinfo {author} {\bibfnamefont{S.~K.}\ \bibnamefont{Bose}},\ }%
\bibfield{journal}{%
	\Doi{10.1063/1.531163}{\bibinfo {journal} {Journal of Mathematical Physics}}\
}%
\textbf{\bibinfo {volume} {36}},\ \bibinfo {pages} {875} (\bibinfo {year}
{1995})\BibitemShut{NoStop}%
\bibitem{Grigore96}%
\BibitemOpen
\bibfield{author}{%
	\bibinfo {author} {\bibfnamefont{D.~R.}\ \bibnamefont{Grigore}},\ }%
\bibfield{journal}{%
	\Doi{10.1063/1.531402}{\bibinfo {journal} {Journal of Mathematical Physics}}\
}%
\textbf{\bibinfo {volume} {37}},\ \bibinfo {pages} {460} (\bibinfo {year}
{1996})\BibitemShut{NoStop}%
\bibitem{Bargmann54}%
\BibitemOpen
\bibfield{author}{%
	\bibinfo {author} {\bibfnamefont{Valentine}\ \bibnamefont{Bargmann}},\ }%
\bibfield{journal}{%
	\Doi{10.2307/1969831}{\bibinfo {journal} {Annals of Mathematics}}\ }%
\textbf{\bibinfo {volume} {59}},\ \bibinfo {pages} {1} (\bibinfo {year}
{1954})\BibitemShut{NoStop}%
\bibitem{Weyl31}%
\BibitemOpen
\bibfield{author}{%
	\bibinfo {author} {\bibfnamefont{Herman}\ \bibnamefont{Weyl}},\ }%
\emph{\bibinfo {title} {The thoery of groups and quantum mechanics}}\
(\bibinfo {publisher} {London},\ \bibinfo {year} {1931})\BibitemShut{NoStop}%
\bibitem{Wigner59}%
\BibitemOpen
\bibfield{author}{%
	\bibinfo {author} {\bibfnamefont{Eugene}\ \bibnamefont{P.Wigner}},\ }%
\emph{\bibinfo {title} {Group Theory and Its Application to the Quantum
		Theory of Atomic Spectra}}\ (\bibinfo {publisher} {Academic Press Inc., New
	York},\ \bibinfo {year} {1959})\BibitemShut{NoStop}%
\bibitem{Ulf63}%
\BibitemOpen
\bibfield{author}{%
	\bibinfo {author} {\bibfnamefont{Ulf}\ \bibnamefont{Uhlhorn}},\ }%
\bibfield{title}{%
	\enquote{\bibinfo {title} {Representation of symmetry transformations in
			quantum mechanics},}\ }%
\bibfield{journal}{%
	\bibinfo {journal} {Arkiv for Fysik}\ }%
\textbf{\bibinfo {volume} {30}},\ \bibinfo {pages} {307} (\bibinfo {year}
{1963})\BibitemShut{NoStop}%
\bibitem{Bargmann64}%
\BibitemOpen
\bibfield{author}{%
	\bibinfo {author} {\bibfnamefont{Valentine}\ \bibnamefont{Bargmann}},\ }%
\bibfield{journal}{%
	\Doi{10.1063/1.1704188}{\bibinfo {journal} {Journal of Mathematical
			Physics}}\ }%
\textbf{\bibinfo {volume} {5}},\ \bibinfo {pages} {862} (\bibinfo {year}
{1964})\BibitemShut{NoStop}%
\bibitem{Note1}%
\BibitemOpen
\bibinfo {note} {Let $\protect \mathcal U_r$ be a continuous ray
	representations of a group $G$. For all $r$ in a suitably chosen neighborhood
	$\protect \mathscr {N}_0$ of the identity $e$ of $G$, one may select a
	strongly continuous set of representatives $U_r \in \protect \mathcal U_r$.
	\par A set of such representatives operator will be called an {\protect \em
		admissible} set of representatives. \cite {Bargmann54}}\BibitemShut{NoStop}%
\bibitem{Note2}%
\BibitemOpen
\bibinfo {note} {From the study of the full Galilean group in $(3+1)$
	dimensions, it is possible to obtain the Bargmann's super-selection rules
	\cite {Brennich70}: \par 1) To different masses there correspond inequivalent
	multipliers and hence inequivalent ray representations; \par 2) ${\protect
		\rm SO} (3)$ has two kind of inequivalent multipliers; one for integer spin
	(unitary representations), a second kind for semi-integer spin (ray
	representations) \cite {Wigner39, GKWigner68}.}\BibitemShut{Stop}%
\bibitem{Note3}%
\BibitemOpen
\bibinfo {note} {Let be $\psi (x,t)$ are the solutions of {Schr\"odinger}
	equation, then: \begin {displaymath} \psi (x,t) \sim (2\pi )^{-\protect \frac
		{3}{3}} \DOTSI \intop \ilimits@ \phi (p) \protect \operatorname e^{-i
		\protect \frac {1}{2 \gamma } \left < {p \protect \tmspace +\thinmuskip
			{.1667em}| \protect \tmspace +\thinmuskip {.1667em} p} \right >t - \left < {p
			\protect \tmspace +\thinmuskip {.1667em}| \protect \tmspace +\thinmuskip
			{.1667em} x} \right > \protect \operatorname d^3 p} \end
	{displaymath}}\BibitemShut{NoStop}%
\bibitem{Messiah75}%
\BibitemOpen
\bibfield{author}{%
	\bibinfo {author} {\bibfnamefont{A.}~\bibnamefont{Messiah}},\ }%
\emph{\bibinfo {title} {Quantum Mechanics, vol.1}}\ (\bibinfo {publisher}
{North-Holland Publishing Company, Amsterdam},\ \bibinfo {year}
{1975})\BibitemShut{NoStop}%
\bibitem{Note4}%
\BibitemOpen
\bibinfo {note} {Where $a_3$ is a real number connected to the Casimir
	element \begin {displaymath} C_3 = 2 H Z_1 - 2 K Z_2 - P^2 \end {displaymath}
	where $H$, $K$, $P$ are time translations, nonrelativistic boosts and space
	translations generators, and $Z_1$ and $Z_2$ are the two central
	elements}\BibitemShut{NoStop}%
\bibitem{Note5}%
\BibitemOpen
\bibinfo {note} {The infinitesimal generators of (\ref
	{eqGG-galilei2Dschrodingernonabelianrep}) are: \begin {align*} H = & \protect
	\tmspace +\thinmuskip {.1667em} \protect \frac {1}{2 \gamma } \left (
	p_1^2+p_2^2 \right ) & N_1 = & \protect \tmspace +\thinmuskip {.1667em}
	\gamma \protect \frac {\partial }{\partial p_1} + i \protect \frac {\lambda
	}{2 \gamma } p_2\\ M = & \protect \tmspace +\thinmuskip {.1667em} is + p_2
	\protect \frac {\partial }{\partial p_1} - p_1 \protect \frac {\partial
	}{\partial p_2} & N_2 = \protect \tmspace +\thinmuskip {.1667em} & \gamma
	\protect \frac {\partial }{\partial p_2} - i \protect \frac {\lambda }{2
		\gamma } p_1\\ P_i = \protect \tmspace +\thinmuskip {.1667em} & p_i \protect
	\tmspace +\thinmuskip {.1667em} (i=1,2) &\protect \tmspace +\thinmuskip
	{.1667em}&\protect \tmspace +\thinmuskip {.1667em} \end {align*} The
	infinitesimal generators of (\ref {eqGG-galilei2Dschrodingerrep}) are: \begin
	{eqnarray*} H &=& \protect \frac {1}{2 \gamma } \left ( p_1^2+p_2^2 \right ),
	\hskip 1em\relax P_i = p_i, \hskip 1em\relax M = is + p_2 \protect \frac
	{\partial }{\partial p_1} - p_1 \protect \frac {\partial }{\partial
		p_2},\\N_i &=& \gamma \protect \frac {\partial }{\partial p_i} (i=1,2) \end
	{eqnarray*} These ones are equivalent also for (3+1)-dimensions
	case.}\BibitemShut{Stop}%
\bibitem{Wigner39}%
\BibitemOpen
\bibfield{author}{%
	\bibinfo {author} {\bibfnamefont{Eugene~P.}\ \bibnamefont{Wigner}},\ }%
\bibfield{journal}{%
	\Doi{10.2307/1968551}{\bibinfo {journal} {Annals of Mathematics}}\ }%
\textbf{\bibinfo {volume} {40}},\ \bibinfo {pages} {149} (\bibinfo {year}
{1939})\BibitemShut{NoStop}%
\bibitem{GKWigner68}%
\BibitemOpen
\bibfield{author}{%
	\bibinfo {author} {\bibfnamefont{E.~P.~Wigner}\
		\bibnamefont{G.~C.~Hegerfeldt}, \bibfnamefont{K.~Kraus}},\ }%
\bibfield{journal}{%
	\Doi{10.1063/1.1664539}{\bibinfo {journal} {Journal of Mathematical
			Physics}}\ }%
\textbf{\bibinfo {volume} {9}},\ \bibinfo {pages} {2029} (\bibinfo {year}
{1968})\BibitemShut{NoStop}%
\bibitem{Brennich70}%
\BibitemOpen
\bibfield{author}{%
	\bibinfo {author} {\bibfnamefont{R.~H.}\ \bibnamefont{Brennich}},\ }%
\bibfield{title}{%
	\enquote{\bibinfo {title} {The irreducible ray representations of the full
			inhomogeneous galilei group},}\ }%
\bibfield{journal}{%
	\bibinfo {journal} {Annals of Institute Henri Poincare'}\ }%
\textbf{\bibinfo {volume} {13}},\ \bibinfo {pages} {137} (\bibinfo {year}
{1970})\BibitemShut{NoStop}%
\end{thebibliography}
\providecommand{\noopsort}[1]{}\providecommand{\singleletter}[1]{#1}%
\end{document}